\documentclass[letterpaper,twocolumn,pra,aps,showpacs,floatfix,superscriptaddress,longbibliography]{revtex4-1}
\usepackage[utf8]{inputenc}
\usepackage{bm}
\usepackage{multirow}
\usepackage{amssymb}
\usepackage{amsbsy}
\usepackage{amsmath,  amsthm, amsfonts, mathrsfs}
\usepackage{graphicx}
\usepackage{epsfig}
\usepackage{placeins}
\usepackage{physics}
\usepackage{trfsigns}
\usepackage[usenames,dvipsnames]{color}
\usepackage[colorlinks,linkcolor=blue,citecolor=blue,urlcolor=blue,breaklinks=true]{hyperref}

\makeatletter

\begin{document} 

 \title{Decoherence Entails Exponential Forgetting in Systems Complying with the Eigenstate Thermalization Hypothesis}

 \author{Lars Knipschild}
 \email{lknipschild@uos.de}
 \affiliation{Department of Physics, University of Osnabr\"uck, D-49069 Osnabr\"uck, Germany}

 \author{Jochen Gemmer}
 \email{jgemmer@uos.de}
 \affiliation{Department of Physics, University of Osnabr\"uck, D-49069 Osnabr\"uck, Germany}

\begin{abstract}
According to the eigenstate thermalization ansatz, matrices representing generic few body observables take on a specific form when displayed in the 
eigenbasis of a chaotic Hamiltonian. We examine the effect of environmental induced decoherence on the dynamics of observables that 
conform with said eigenstate thermalization ansatz. The obtained  result refers to a description of the dynamics in terms of an integro-differential 
equation of motion of the Nakajima-Zwanzig form. We find that environmental 
decoherence is equivalent to an exponential damping of the respective memory kernel. This statement is formulated as rigorous theorem. Furthermore the 
implications of the theorem on the stability of exponential dynamics against decoherence and the transition towards Zeno-Freezing are discussed.

\end{abstract}


\maketitle

\section{Introduction}
Coupling to some  environment drives the local density matrix of a quantum system towards a diagonal form in a specific basis - this well known 
finding is at the heart of open quantum system theory \cite{breuer}. If this influence is such that its  sole effect is to erase off-diagonal 
elements in the specific basis but leave diagonal elements unchanged, the process is sometimes called pure decoherence or pure dephasing 
\cite{classical_world,Zurek_Decoherence, Skinner_Decoherence,Alicki_Decoherence}. 
The (not necessarily orthogonal)
basis states of the eventually diagonal 
density matrix correspond to  ``pointer states''. They are singled out by ``environment induced superselection'' and depend strongly on the observables through which a 
system couples to its environment. Some sort of decoherence is almost inevitably induced by any complex environment \cite{Esposito_Relaxation},
specific pointer states come with 
environments that may be thought of as monitoring some system observable, the pointer states then essentially being the eigenstates of the monitored observable. In both cases the 
decoherence process is routinely modeled by corresponding quantum master equations, often of Lindblad form \cite{breuer}. Environmental decoherence may in 
general alter the dynamics of any system observable substantially.

Somewhat more recent but similarly intensely debated is the eigenstate thermalization hypothesis (ETH) \cite{Deutsch_ETH,srednicki99,Rigol_Thermalization}. 
Most encompassing the ETH may be 
described as  a statement on the properties of the matrix elements $\langle l| A|m \rangle$ of some observable $A$ when represented in the energy eigenbasis 
$\{|l(m) \rangle\}$. According to the ETH ansatz, the diagonal elements are very similar when corresponding to similar energies, whereas the off-diagonal elements 
closely resemble a set of independent Gaussian random numbers, with zero mean and variances  that smoothly depend on the position of the matrix element within the matrix.
While rigorous conditions under which the 
ETH ansatz applies are yet unknown, there are plenty of numerical examples which confirm its applicability to standard observables in interacting 
many-body systems \cite{Rigol_ChaosETH_2016,Haque_2015,Rigol_ETH_2017,Rigol_ChaosETH_2016}. Generally validity of the ETH is expected for few-body observables in 
non-integrable systems. As one  consequence of the ETH, expectation values effectively 
(up to  Poincare recurrences) dynamically relax towards their equilibrium values as calculated from the respective ensembles 
(canonical, microcanonical, generalized  Gibbs, etc.) The ETH, however, also entails a certain property of the actual relaxation behavior itself: It appears that 
if the ETH ansatz applies, a common relaxation behavior of the expectation value results,
for a multitude of initial states $\ket{\psi_k(0)}$, i.e. $\langle \psi_k(0) |A(t) \psi_k(0) \rangle \approx  \langle \psi_k(0) |A(0)\psi_k(0) \rangle g(t)$. 
($g(t)$ is essentially given by the respective 
correlation function $g(t)\propto \langle A(t)A\rangle$) \cite{srednicki99,Richter}. This statement includes also and especially initial states far from 
equilibrium. While the statement itself and the concrete range of its validity are currently  under scrutiny, we somewhat boldly move ahead with this this work 
and focus on the principles
that arise if the validity is taken for granted (for a class of states to be defined below) and combined with the decoherence due to an environment. (To support 
the above ETH statement in a non-rigorous way, we simply provide some evidence for its validity based on numerical analysis of pertinent examples, 
see Appendix \ref{evidence}.)

In the paper at hand we thus investigate the influence of a decohering environment on the dynamics of an observable in a system that, as an isolated system, fulfills the 
ETH. We present a theorem which establishes 
that dynamical decoherence is then 
strictly equivalent to  an exponential damping of the memory kernel, if $\langle A(t)\rangle$ is described by an integro-differential equation of motion
such as a pertinent Nakajima-Zwanzig equation \cite{kubo2012statistical}. To rephrase in an informal manner, 
the stronger the decoherence is, the quicker the system ``forgets'', yielding   a more Markovian behavior.

\section{formal statement of the main theorem}

The main result  of the present work may be formulated in terms of a theorem which we formally present in the following. Generally, we consider  a quantum system S
the dynamics of which are (effectively) restricted to a finite, $N$-dimensional Hilbert space $\mathcal{H}$.
Let $A = \sum_{j=1}^n a_j \ket{j}\bra{j}$ be an Hermitian operator on $\mathcal{H}$. For simplicity we assume $A$ to be non-degenerate, i.e., $a_j \neq a_k$ if $j \neq k$.
(This assumption may be dropped, but it clarifies the presentation substantially and appears natural for S being chaotic, which will be assumed below,
see \textit{Condition 3}.)
Then $A$ entails an unique, complete, orthonormal basis $\mathfrak{B}$ of
$\mathcal{H}$ composed of its eigenvectors, i.e.,  $\mathfrak{B} = \{ \ket{1}, ...,\ket{j},... ,\ket{N} \}$.
The operator $A$ represents the observable of interest. Denote furthermore the  operator representing the 
Hamiltonian proper on S by $H$ and the density operator of S by $\rho$.
Below S will be treated as either a closed or an open system, depending on the respective pertinent equation of motion, see \textit{Condition 1}.
 
\textit{Condition 1: Decohering Dynamics }\\
Let the dynamics of $\rho$ be generated by the following quantum master equation:
\begin{equation}
    \label{lindblad}
    \begin{aligned}
    & \frac{\partial \rho}{\partial t} = -i [H, \rho] + \frac{\gamma}{2} \left( \sum_{j=1}^n 2 L_j \rho L_j^\dagger - L_j^\dagger L_j \rho - \rho L_j^\dagger L_j \right) \\
    & L_j:=\ket{j}\bra{j}
    \end{aligned}
\end{equation}
(throughout the paper we tacitly set $\hbar = 1$). At $\gamma = 0$ the environment is decoupled and $\rho(t)$ follows the respective closed system dynamics. 
At  $\gamma > 0$, (\ref{lindblad})
is of the standard (Lindblad) form which is routinely used to model the influence of a weakly coupled environment, the only effect of 
which is to (effectively) continuously measure $A$. For simplicity 
an ``infinite resolution'' of the environment is assumed, i.e, any two projective eigenspaces of $A $ decohere equally quickly, i.e, with the rate $\gamma$. While this 
``uniform decoherence'' does not necessarily occur (it is, e.g., strongly violated in the Caldeira-Leggett model with respect to position) it serves here as a convenient 
starting point capturing the essential physics. However, other concrete models featuring uniform decoherence are discussed, e.g., in Refs. 
\cite{Esposito_Diffusion,classical_world,znidaric10,znidaric13,Kassal_Transport}

\textit{Condition 2: Eigenstate Thermalization Hypothesis}\\
Let $ A(t) $ denote the Heisenberg representation of $A$ with respect to the dynamics of the isolated system S. Then we require the following equation to hold:
 
\begin{equation}
    \label{richter}
    \sum_{j=1}^n p_j \bra{j} A(t) \ket{j} = \left( \sum_{n=1}^n p_j \bra{j} A \ket{j} \right) g(t)
\end{equation}
for all $p_j$. $g(t)$ is a real-valued function of time. Eq. (\ref{richter}) implies that, up to a prefactor, the evolution of the expectation value of $A$ is always the 
same, if the initial state is any eigenstate of $A$, irrespective of the particular eigenstate. While this is a strong condition, there is evidence that it may be 
fulfilled to remarkable accuracy, if $A$ complies with the ETH ansatz \textcolor{black}{as given, e.g., in \cite{srednicki99}. (For ``self-containedness'' we elaborate on 
the ETH ansatz in Appendix \ref{evidence}). Note that the ETH ansatz is a statement on all matrix elements of $A$ as represented in the eigenbasis of $H$.}
The above evidence includes analytical reasoning 
as well as numerical examples based on spin systems
\cite{Richter,srednicki99,Khatami}. In Appendix \ref{evidence} we provide more evidence based on partially random matrices 
in accord with the ETH ansatz.

\textit{Condition 3: Diagonal Initial State}\\ 
Let the initial state $\rho_0$ be of the following form:
\begin{equation}
    \label{init_state}
    \rho_0 = \sum_{j=1}^n c_j \ket{j}\bra{j}, \; \; c_j \ge 0, \; \; \sum_{j=1}^n c_j    =1.
\end{equation} 
This simply restricts the possible set of initial sates to those that are diagonal with respect to the above specific eigenbasis of $A$.

{\it Definition: Memory-Kernel}\\
Let the the memory-kernel $\kappa (\tau)$ corresponding to a function $\alpha(t)$, be implicitly defined by the following expression:
\begin{equation}
\label{mk_def}
\dv{\alpha (t)}{t} = -\int_0^t \kappa(t-t') \alpha(t') \mathrm{d} t' = - \kappa * \alpha(t)
\end{equation}
Note that (\ref{mk_def}) establishes a bijective map between the functions $\alpha(t)$ themselves and their respective memory kernels $\kappa (\tau)$ together with
the initial values of the functions, $\alpha(0)$. To rephrase:  $\kappa (\tau)$ may be calculated from $\alpha(t)$; knowledge of $\kappa (\tau)$ and  $\alpha(0)$ suffices 
to calculate $\alpha(t)$. This bijectivity plays a pivotal role in the derivation of the theorem.
Note that (\ref{mk_def}) is not a condition or 
assumption, it is simply a definition which is applicable to all Laplace-transformable functions $\alpha(t),\kappa (\tau)$.

\textit{Theorem:} Let $a(t)$ and $\tilde{a}(t)$ be the dynamical expectation values of $A$, i.e. $\langle A(t) \rangle = \mathrm{Tr}\{A(t)\rho_0\}$, 
without ($\gamma=0$) and with ($\gamma > 0$) the influence 
of the environment, respectively.
Let $K(\tau)$ and $\tilde{K}(t)$ be the respective  Memory-Kernels according to (\ref{mk_def})). To rephrase, let the following attribution apply: 
$\alpha(t) = a(t) \Leftrightarrow  \kappa(t)= K(t)$ and, respectively, $\alpha(t) = \tilde{a}(t) \Leftrightarrow \kappa(t)= \tilde{K}(t)$

If the conditions (\ref{lindblad}), (\ref{richter}) and (\ref{init_state}) are met, the influence of the environment on the dynamics of 
the expectation value may strictly and completely be  captured  in the following equation:

\begin{equation}
    \label{main_res}
    \tilde{K}(\tau) = K(\tau) \exp(-\gamma \tau)
\end{equation}
This finding is our main result. An explicit proof of the theorem is given in Appendix \ref{proof}.

Before discussing the main  result from a more general perspective in the concluding paragraphs, we proceed by outlining a possible scheme of 
application of the theorem. Apart from its practical relevance, this scheme is intended to convey the essence of (\ref{main_res}) most clearly.
Furthermore we establish the direct implication of the theorem on the stability of exponential decay dynamics and 
on the behavior in the strong decoherence regime, i.e., at the 
transition to Zeno-freezing.

 \section{implications of the main theorem}

\textit{ Scheme of Application of the Theorem:} \\
Relation (\ref{main_res}) allows for the computation of the expectation value dynamics under the influence of the environment, $\tilde{a}(t)$, from the 
respective dynamics of the isolated system, $a(t)$, without the need to solve the Lindblad equation. This is may be done  by applying  the following scheme:
\begin{equation}
    a(t) \xrightarrow{(\ref{mk_def})} K(\tau) \xrightarrow{(\ref{main_res})} \tilde{K}(\tau) \xrightarrow{(\ref{mk_def})} \tilde{a}(t)
\end{equation}
First one  computes the memory-kernel $K(\tau)$ from $a(t)$ using (\ref{mk_def}). Next theorem (\ref{main_res}) is employed to calculate the 
memory-kernel $\tilde{K}(\tau)$. Eventually (\ref{mk_def}) is used  again to calculate the perturbed dynamics $\tilde{a}(t)$ from $\tilde{K}(\tau)$.

\textit{Corollary 1: Stability of exponential decay}\\
 Let  $a(t)$ as resulting from (\ref{lindblad}) without the influence of the environment ($\gamma =0$) be strictly exponential with some decay constant $\beta$, i.e.
 
 \begin{equation}
    \label{exponential1}
    a(t) = a(0) \exp(-\beta t).
\end{equation}
(While dynamics strictly according to (\ref{exponential1}) are impossible in in any finite system, numerous examples exist in which the expectation value dynamics 
are very well approximated by  (\ref{exponential1}).) Then the dynamics of the expectation value under decohering influence of the environment is given by

\begin{equation}
    \label{exponential2}
   \tilde{a}(t) =  a(t) = a(0) \exp(-\beta t),
\end{equation}
meaning the dynamics remains unaltered.
Eq. (\ref{exponential2}) is readily inferred from 
(\ref{main_res}): The memory kernel $K(\tau)$ corresponding to $a(t)$ as given in (\ref{exponential1}) is  $K(\tau)= \beta \delta (\tau)$. Thus, for dynamical 
decoherence of strength $\gamma$, the respective memory kernel  $\tilde{K}(\tau)$ is given by
\begin{equation}
    \label{exponential3}
   \tilde{K}(\tau)=\beta \delta (\tau) \exp(-\gamma \tau) =\beta \delta (\tau)=K(\tau).
\end{equation}
Since, according to (\ref{mk_def}), equal memory kernels imply equal dynamics of the expectation values, (\ref{exponential2}) follows directly. 

\textcolor{black}{As already mentioned below (\ref{exponential1}) neither ``kinks'' in the observable dynamics nor $\delta$-functions in the respective memory-kernels can 
truly appear in finite systems. Thus Corollary 1 describes simply the  limiting case in which the  decay of the memory kernel is much shorter than 
the decohering dynamics induced by the environment. More specifically: the decay of the memory kernel must be shorter than $\gamma^{-1}$. 
The state of affairs outside this regime is discussed below in Corollary 2.}

Note that, given the 
validity of (\ref{main_res}), the exponential relaxation behavior (\ref{exponential1}) is the only stable form of relaxation. All other forms will inevitably be 
affected by decoherence.

\textit{Corollary 2: Transition to Zeno-Freezing}\\
Let the memory kernel $K(\tau)$ corresponding to the isolated dynamics   $a(t)$ be a non-singular, analytic function at $\tau =0$ 
(Note that this assumption excludes memory kernels as addressed by the previous corollary, c.f. (\ref{exponential3})). Then, for sufficiently strong 
decorherence, i.e.,  $\gamma \geq \gamma'$, the dynamics  $\tilde{a}(t)$ take the form

\begin{equation}
    \label{zeno1}
   \tilde{a}(t) \approx   a(0) \exp(-\frac{K(0)}{\gamma} t).
\end{equation}
Thus, strong decoherence first renders the dynamics exponential, and then slows it down until it freezes. This also follows directly from (\ref{main_res}). Above some 
 $\gamma \geq \gamma'$, the following approximation for the memory kernel of the decohering  dynamics   $\tilde{K}(\tau)$ (given the above analyticity) holds:
 
\begin{equation}
    \label{zeno2}
  \tilde{K}(\tau) = K(\tau) \exp(-\gamma \tau) \approx  K(0) \exp(-\gamma \tau) 
\end{equation}
If furthermore $\gamma^2 \gg K(0)$, the decay of the dynamics  $\tilde{a}(t)$ is much slower than the decay of the corresponding memory kernel
 $\tilde{K}(\tau)$. Thus, timescale separation may be applied and the validity of (\ref{zeno1}) {\it a posteriori} inferred.  
 
To clarify all aspects of (\ref{zeno1}) we note that the initial value of the memory-kernel may be found rather easily. According to, e.g., the Mori memory-matrix 
formalism it is:
\begin{equation}
    \label{mori}
  \tilde{K}(0) = \frac{\text{Tr}\{ A [H,[H, A]]   \}}{\text{Tr}\{ A^2    \}}.
\end{equation}

\section{Physical Discussion of the Results}
We now embark on a discussion of the above findings.

Environment induced decoherence is omnipresent \cite{classical_world}. The ambitious attempt to get rid of all environmental decoherence is one of the driving forces 
behind 
the research on ultra-cold atoms at present \cite{Gross_ColdAtoms}. However,  for many less perfectly isolated set-ups, the timescale on  which
the coherence between, say, 
energy eigenstates vanishes is much shorter than the timescale on which substantial amounts of energy are exchanged with the environment 
\cite{breuer, weiss2012quantum}. 
Hence, in principle, the presence of the 
environment may very well have significant influence on the dynamics of local observables, even at times at which practically no energy has been exchanged yet. This 
also applies to cases in which the system S itself is rather large or even macroscopic. Consider now, as a cartoon example, two macroscopic objects,
1 and 2, which may exchange energy with each other, but are energetically insulated from the rest of the world. Nonetheless, the set up  inevitably  comprises 
some environment, 
e.g. some styrofoam box, etc. Take the energy difference $A:= H_1-H_2$ as the observable of interest. As the environment ``senses'' the local energies quickly, an 
equation of motion similar to Eq. (\ref{lindblad}) is adequate to model this scenario at relevant timescales. Thus, to repeat, the
decohering influence could in principle very well have 
substantial impact 
on the concrete dynamics of $\langle A(t)\rangle$. Such an influence, however, is undeniably never observed. The concrete nature of the environment, such as styrofoam, 
insulation by 
partial vacuum, etc. is evidently entirely irrelevant for the evolution of  $\langle A(t)\rangle$. This independence may be explained based on 
\textit{Corollary 1}: As long $\langle A(t)\rangle$ relaxes exponentially, the independence of the strength of the
decoherence is accounted for by theorem (\ref{main_res}).
This statement challenges the paradigm of exponential, Markovian dynamics always being due to the decohering influence of some
environment. Recall that $\langle A(t)\rangle$ is the evolution of the observable without the environment. In spite of this paradigm there are, however, 
examples of exponentially relaxing (macroscopic) observables in finite  quantum systems that are entirely isolated, i.e. not coupled to any bath 
\cite{Steinigeweg_Diffusion_Hubbard,Niemeyer_Spin_Thermalization,Niemeyer_Fokker_Planck}. 
The above reasoning strongly indicates that all macroscopic, exponential relaxation scenarios must be of this type. If the exponential relaxation was induced by the 
environment, the above independence would be absent, and relaxation dynamics would change with different decohering environments and respective couplings.
An analogous argument is suitable to explain the stability of the dynamics of heat-exchange in and between macroscopic objects against environmental 
decoherence in general. This reasoning is conceptually in line with Ref. \cite{Monteoliva_Decoherence} where considerable effort goes to demonstrate that local entropy production is 
insensitive to decoherence strength on a wide regime.

The essence of \textit{Corollary 2} may be captured conveniently from
comparing it, e.g., to Refs. \cite{Esposito_Diffusion,znidaric13}: In these works
the respective authors
elaborate on  the effect of
spatial decoherence on
the transport properties of particle(s)  in ordered \cite{Esposito_Diffusion} as
well as disordered \cite{znidaric13} tight-binding lattices.
They find that decoherence induces a gradual transition from ballistic
(ordered) or localized (disordered) to
diffusive
dynamics. Diffusive dynamics may be described in terms of Markovian
random walks, i.e., exponential decay of spatial density waves.  Such a
description does not apply to
ballistic or localized  dynamics. Thus, Refs. \cite{Esposito_Diffusion,znidaric13}
describe the decoherence induced transition  towards exponential
dynamics of  a macrovariable.
Furthermore,
increasing decoherence
strength is shown to slow the dynamics down until it freezes. The
results of both works  \cite{Esposito_Diffusion,znidaric13} are in quantitative
accord with  \textit{Corollary 2}.
The latter, however, establishes such a transition  for all scenarios
that may generally described as ETH-conforming systems in
(strongly) decohering environments.

\section{summary and conclusion}
We analyzed systems in which
some observable exhibits dynamics in accord with the eigenstate
thermalization hypothesis.
While this condition does not restrict the functional form of the
short time dynamics very severely (in contrary to the long time
dynamics), it qualitatively
requires the short time dynamics to be very similar to each other (up to
a corresponding prefactor) for a large class of different initial
states. If such a system is
put into contact with an environment that may be described as
``measuring'' the respective observable, the influence of the past on
the temporal change of the
observable at present, gets exponentially damped. This statement has been
delivered as a rigorous theorem. One direct consequence of the theorem
is the stability of Markovian
observable dynamics (which feature short memory anyway) against
decoherence, and thus varying decoherence strengths. This contributes
to  an understanding of the
factual independence of all sorts of Markovian macro- and mesoscopic
relaxation processes of the precise nature of their different,
inevitably present,
decohering environments. Furthermore
it may be inferred  that dynamics, that are non-Markovian in the
isolated system, will become simple exponential relaxations at strong
environmental decoherence, close to Zeno-freezing. 
From a practical point of view we suggest a
method to calculate observable dynamics under the influence of an
environment from the
dynamics in the isolated system, without solving any quantum master
equation. This may facilitate the description of dynamics of systems
which live in a regime in between
coherent and incoherent, such as quantum biological systems \cite{Rebentrost_Transport, Plenio_Transport} etc.

This work has been funded by the
Deutsche Forschungsgemeinschaft (DFG) - GE 1657/3-1.
We sincerely thank the members of the
DFG Research Unit FOR 2692 for fruitful discussions.

\bibliography{mybib}

\appendix

\section{Eigenstate Thermalization Ansatz and Evidence Supporting Condition 2.} \label{evidence}

In this Section we provide evidence that the behavior required by \textit{Condition 2} may indeed be expected for systems and observables conforming 
with the eigenstate thermalization ansatz. For comprehensibility we only present numerical examples based on partially random matrices here. The interested
reader may refer to \cite{srednicki99,Khatami,Richter} for corresponding analytical reasoning and spin- as well as boson-based examples.

According to Ref. \cite{srednicki99} the ETH ansatz for the matrix elements ${a}_{jl}$ of the observable $A$ with respect to the energy eigenbasis reads:
\begin{equation} 
{a}_{jl} = {\cal A}(E) \, \delta_{jl} + 
\Omega(E)^{-1/2} \, f(E, \omega) \, R_{jl} \ ,
\label{ETH}
\end{equation}
where $E:=(E_j + E_l)/2,  \quad \omega:= E_j - E_l$ and the $E_{j(l)}$ are the eigenvalues of the respective Hamiltonian.
The density of states is denoted by $\Omega(E)$ and ${\cal A}(E),   f(E, \omega)$ are smooth functions of their 
arguments. And, to cite  \cite{srednicki99}, ``$R_{jl}$ is  a  numerical  factor  that
varies erratically with $j$ and $l$.  It is helpful to think of the real and imaginary parts of $R_{jl}$
as random variables, each with zero mean and unit variance.'' Below we implement this concept  by simply choosing 
the $R_{jl}=R_{lj}  $ as  i.i.d. Gaussian random real numbers  with zero mean and unit variance.

\begin{figure*}[h]
\includegraphics[width=0.49\textwidth]{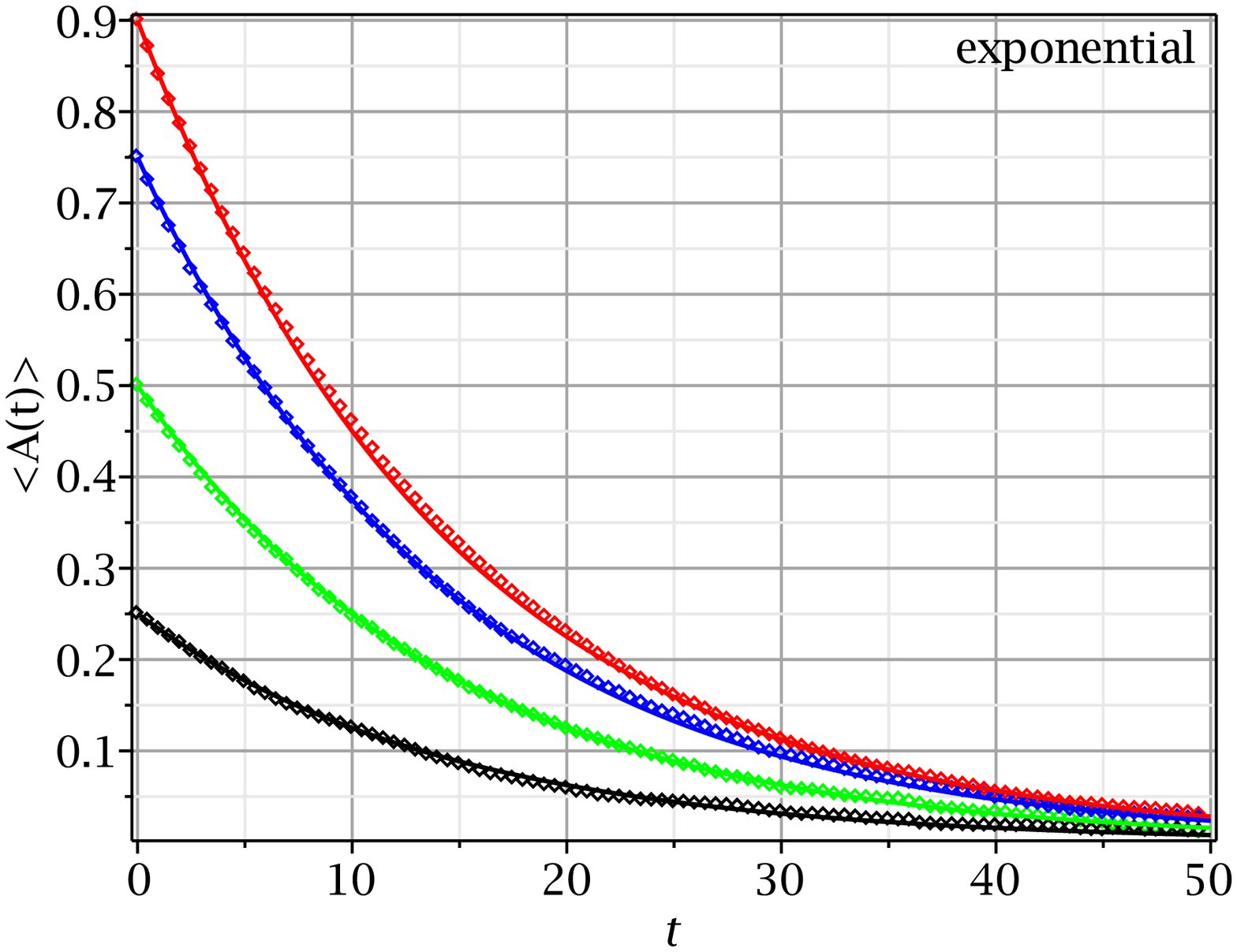} 
\includegraphics[width=0.49\textwidth]{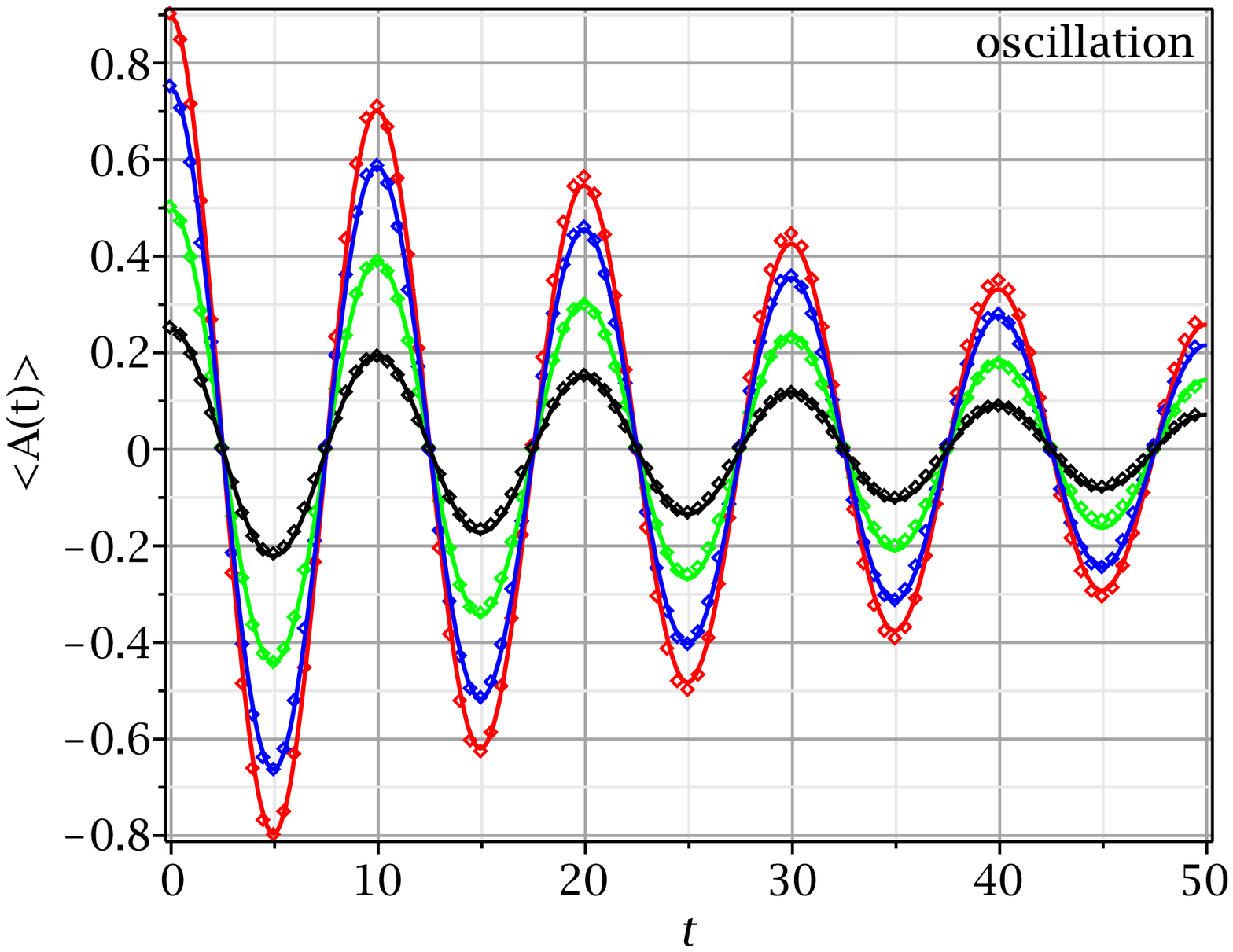}
\includegraphics[width=0.49\textwidth]{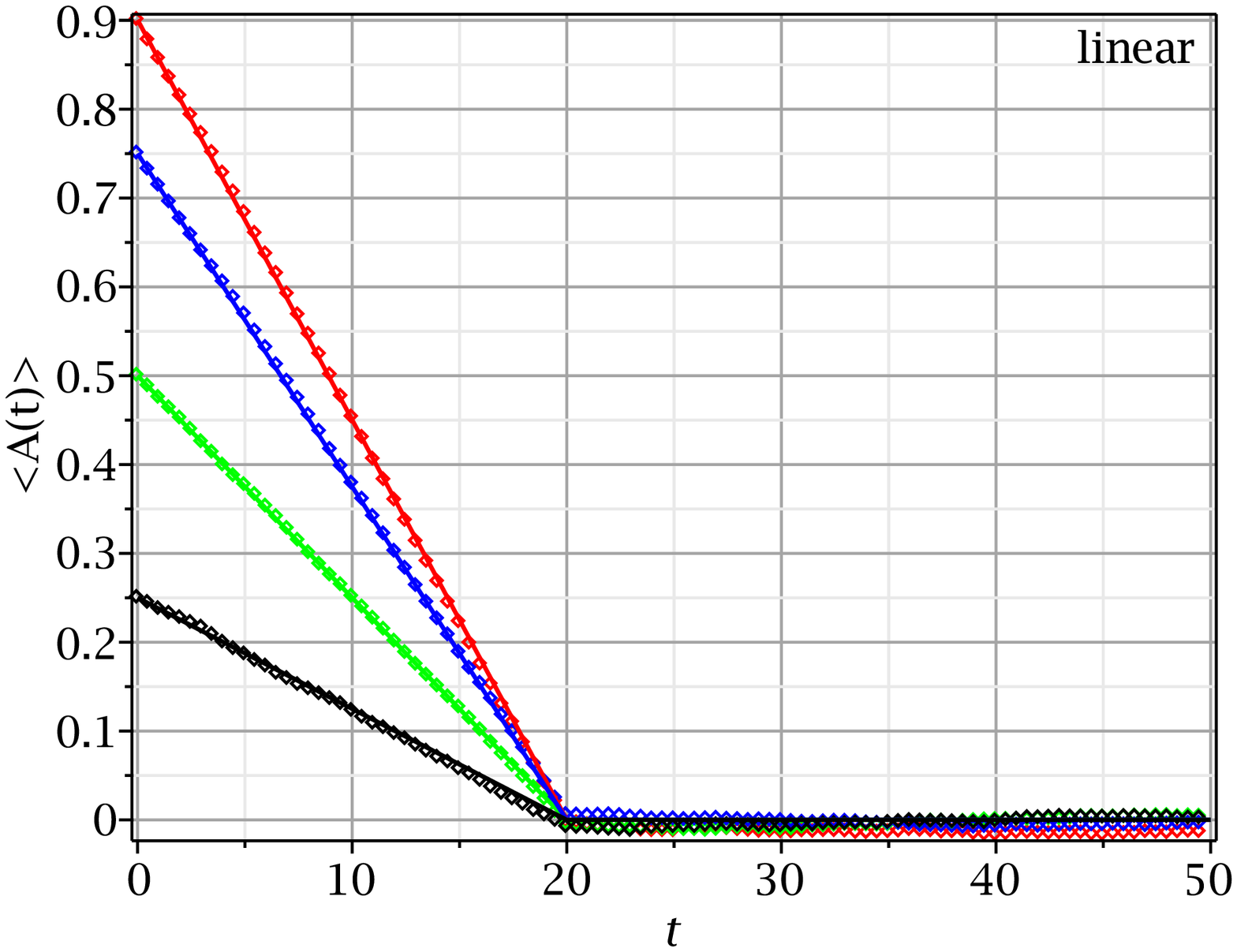}
\includegraphics[width=0.49\textwidth]{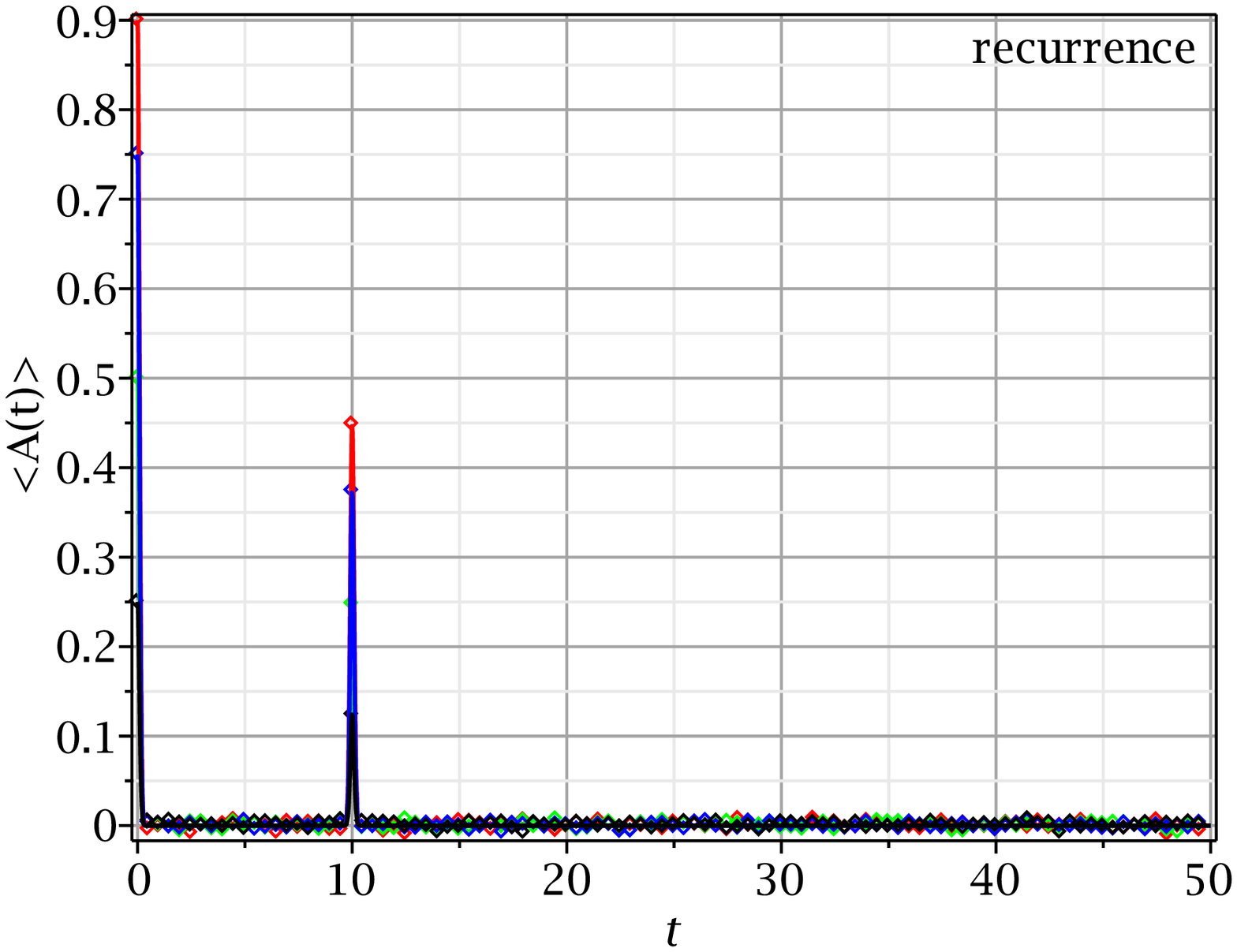} 
\includegraphics{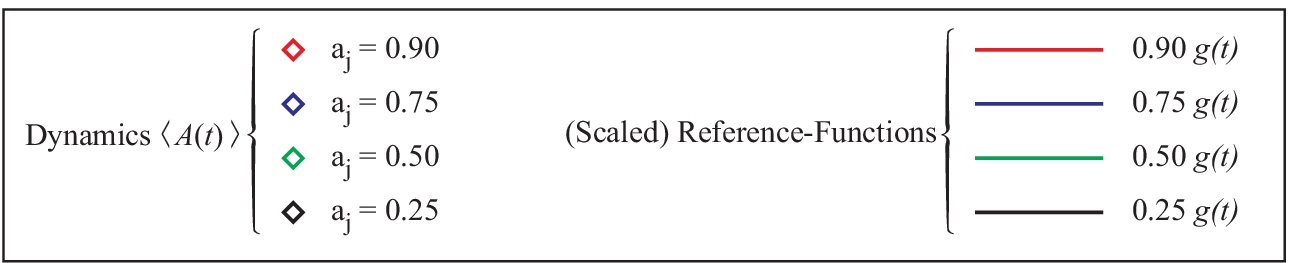} 
\caption{We implemented four pairs of Hamiltonian $H$ and observable $A$ in accord with the ETH ansatz. These implementations are constrained to render $\langle A(t) A \rangle \propto g(t)$ (see Tbl. \ref{ref_functions} for the different $g(t)$).
We calculated the expectation value $\bra{j} A(t) \ket{j}$ for four different eigenstates $\ket{j}$ of $A$ (diamonds).
The dynamics of $\bra{j} A(t) \ket{j}$ are in good accordance with the scaled reference-functions (curves).
}
\label{numerical_results}
\end{figure*}

While rigorous conditions under which the 
ETH ansatz applies are yet unknown, there are plenty of numerical examples which confirm its applicability to few-body observables in non-integrable 
(in the sense of a Bethe ansatz) models of interacting particles on lattices.
\cite{Rigol_ChaosETH_2016, Haque_2015, Rigol_ETH_2017, Rigol_ChaosETH_2016}
For simplicity we choose in all our below examples: ${\cal A}(E)=0, f(E, \omega)$ 
independent of $\omega$, i.e., $f=f(\omega)$ and $\Omega(E)$ independent of $E$, i.e., $\Omega(E)=\Omega$. If the dynamics are restricted to a (narrow) energy shell, it
suffices to model only the sector of $A$ corresponding to that energy shell. Within such a sector the above choices are natural \cite{srednicki99}. We accordingly 
construct the matrix elements as 
\begin{equation}
 a_{jl} = f(\omega) R_{jl}
 \label{eth0}
\end{equation}
Any specific choice of  $f(\omega)$ corresponds to some auto-correlation function Tr$\{A(t)A\}$. If (\ref{richter}) applies,
then this auto-correlation function essentially determines 
the ``universal relaxation function'', i.e., $g(t)\propto$ Tr$\{A(t)A\}$. To limit numerical effort we restrict the analysis to four  
different exemplary $g(t)$ (corresponding to four different $f(\omega)$), see Tab. \ref{ref_functions}. 
For graphs of the functions $g(t)$ see Fig. \ref{numerical_results}.
\begin{table}
\caption{Sample of four exemplary expectation value dynamics $g(t)$, for graphs see Fig. \ref{numerical_results}. The specific forms have been chosen to render
timescales comparable, $\tau = 10.0$.}
\centering
\begin{tabular}{ll}
\hline
\textbf{Name} & \textbf{Definition} \\
\hline
Exponential & $g_{\mathrm{exp}}(t) = \exp(-\frac{\ln{2}}{\tau} |t|)$ \\
Oscillation & $g_{\mathrm{osc}} = \cos(\frac{2 \pi}{\tau} t) \exp(-\frac{1}{2 \tau} |t|)$ \\
Linear & $g_{\mathrm{lin}} = \begin{cases} 
	1-\frac{|t|}{2 \tau} & |t| \le 2 \tau \\
	0 & \mathrm{otherwise}
\end{cases}$ \\
Recurrence & $g_{\mathrm{rec}}(t) =$ $\exp(-\frac{t^2}{v}) + $ \\
& $0.5 \exp(-\frac{(t-\tau)^2}{v}) +$ \\
& $0.5 \exp(-\frac{(t+\tau)^2}{v})$; $v=0.016$ \\
\hline 
\label{ref_functions}
\end{tabular}
\end{table}
The functional form of $g(t)$ determines the respective $f(\omega)$ only up to a pre-factor. We fix this prefactor such as to render the largest eigenvalue of $A$, 
$a_\text{max}$ equal to unity, i.e., $a_\text{max}\stackrel{!}{=} 1$.
The $g(t)$'s corresponding to exponential decay and  damped oscillation have been  chosen to represent
standard forms of equilibration dynamics which are known to occur frequently for a variety of systems. The linear and the recurrence dynamics represent exceptional  
evolutions that are practically hardly ever encountered. We include the latter in our analysis to clarify that the validity of {\it Condition 2} is not restricted to 
standard relaxations such as exponential, etc.
In accord with the above choice  $\Omega(E)=\Omega$ we choose the $N$ eigenvalues $E_{j(l)} $ as 
uniformly i.i.d. random numbers from the interval $[-30,30]$.
Since we are mainly interested in the thermodynamical limit (large $N$) we performed some of the following numerical investigations for different $N$ from 10000 to 70000.
We found the relevant numerical results to be essentially free of finite size effects at or above  $N=20000$, thus the below data have been computed at this dimension. 

To exemplarily check the applicability of  (\ref{richter}) we computed $\bra{j} A(t) \ket{j}$ for four different initial states $\ket{j}$ corresponding to 
$a_j \approx 0.25, 0.5, 0.75, 0.9$ for each decay model. Since Tr$\{A\} = 0$ and $a_\text{max} = 0.9$, these initial states sample almost 
the entire (positive) spectrum of $A$. 
The results are displayed in Fig. \ref{numerical_results}. We plot the expectation value dynamics $\bra{j} A(t) \ket{j}$ for each  type of dynamics. Obviously the 
 $\bra{j} A(t) \ket{j}$ coincide up to a factor $a_j$  to very good accuracy. While this is of course just exemplary numerical evidence, these results strongly suggest the
 validity of (\ref{richter}) for observables complying with the ETH ansatz (\ref{ETH}).

\section{Proof of the main Theorem} \label{proof}
In this section we proof (\ref{main_res}). 
\subsection{Discretization of (\ref{lindblad})}
As a first step we rewrite (\ref{lindblad}) in a time-discrete form.
Since the time-step $T=t/N$ is infinitesimal small, we neglect terms of the order $\mathcal{O}(T^2)$.
\begin{equation}
    \label{lindblad_discrete}
    \rho_n = (1-\gamma T) U \rho_{n-1} U^\dagger + \gamma T \hat{P} U \rho_{n-1} U^\dagger + \mathcal{O}(T^2)
\end{equation}

$U$ denotes the propagator of the closed system for the time-step $T$. $\hat{P}$ is a super-operator that removes all
non-diagonal elements in the Basis $\mathfrak{B}$. $\rho_n$ denotes the density matrix at the time $t=n T$.

\subsection{A recursive model to calculate the expectation value in the decohered system}
We now consider the expectation value of $A$:
\begin{equation}
    \langle A \rangle_{\rho_n} = (1-\gamma T) \tr(A U \rho_{n-1} U^\dagger) + \gamma T \tr(A U \hat{P} \rho_{n-1} U^\dagger)
\end{equation}

We define $z_n(\epsilon)$ recursively:
\begin{equation}
    \label{decoherence_model}
\begin{aligned}
    z_0(\epsilon) = & \langle A \rangle_{\rho_0} g(\epsilon) \\
    z_n(\epsilon) = & (1-\gamma T) z_{n-1}(\epsilon + T) + \gamma T z_{n-1}(0) g(\epsilon + T)
\end{aligned}
\end{equation}

In order to show that
\begin{equation}
    \langle A \rangle_{\rho_n} = z_n(0)
\end{equation}
is valid, we prove the following equation by induction (over $n$):

\begin{equation}
    \label{ind_assumpt}
    \tr(A U^m \rho_n (U^\dagger)^m) = z_n(mT); m, n \in \mathbb{N}
\end{equation}

The base-case ($n=0$) directly follows from (\ref{richter}), since the initial state $\rho_0$ is diagonal by definition:
\begin{equation}
    \tr(A U^m \rho_0 (U^\dagger)^m) = \langle A(m T) \rangle_{\rho_0} = \langle A \rangle_{\rho_0} g(m T) = z_0(m T)
\end{equation}

We now prove that, if (\ref{ind_assumpt}) holds for all $n \in \{0, 1, ..., k-1\}$ with $k \in \mathbb{N}$, it holds for $n=k$, as well.
\begin{equation}
    \label{}
    \begin{aligned}
        & \tr(A U^m \rho_k (U^\dagger)^m) \\
        =& (1-\gamma T) \tr(A U^{m+1} \rho_{k-1} (U^\dagger)^{m+1}) \\
        +& \gamma T \tr(A U^{m+1} \hat{P} \rho_{k-1} (U^\dagger)^{m+1})
    \end{aligned}
\end{equation}
We applied the the discrete form of the Lindblad-equation (\ref{lindblad_discrete}).
The first summand in this equation can be simplified by applying the induction-assumption (\ref{ind_assumpt}).
Since $\hat{P} \rho_{k-1}$ exhibits diagonal-form regarding $\mathfrak{B}$ we use (\ref{richter}) to simplify the second summand.
After applying the induction-assumption (\ref{ind_assumpt}), we finally find:
\begin{equation}
    \begin{aligned}
        & \tr(A U^m \rho_k (U^\dagger)^m) \\
        =& (1-\gamma T) z_{k-1} ((m+1) T) + \gamma T z_{k-1}(0) g((m+1) T)
    \end{aligned}
\end{equation}

\subsection{Transforming the recursive model into a partial differential equation}
Since the time-step $T$ is infinitesimal, we now transform the recursive definition of $z_n$ into a partial differential equation (pde).
Therefore we define $z(t=nT,\epsilon) = z_n(\epsilon)$ and rewrite (\ref{decoherence_model}):
\begin{equation}
    \label{discrete_model}
    \begin{aligned}
        z(t,\epsilon) &=(1-\gamma T) z(t-T,\epsilon+T) \\
            &+ \gamma T z(t-T, 0) g(\epsilon + T) 
    \end{aligned}
\end{equation}

We linearly approximate $z$ for small $T$ and find:
\begin{equation}
    \label{pde}
    \begin{aligned}
        \frac{1}{g(\epsilon)} \left( \gamma z(t,\epsilon) + \frac{\partial z}{\partial{t}} - \frac{\partial z}{\partial \epsilon} \right) = \gamma z(t,0)
    \end{aligned}
\end{equation}
For infinitesimal small $T$ this approximation and (\ref{discrete_model}) become equivalent.



The function $z(t, \epsilon)$, that solves this equation and satisfies the boundary condition (\ref{decoherence_model}), reads:
\begin{equation}
    \begin{aligned}
        z(t,\epsilon) &= a(\epsilon+t) \exp(-\gamma t) \\
         &+ \int_0^t \gamma z(t',0) g(t+\epsilon-t') \exp(-\gamma (t-t')) \mathrm{d} t'
    \end{aligned}
\end{equation}
This can be verified by inserting $z(t,\epsilon)$ into (\ref{pde}).

Since we are only interested in the dynamics $\tilde{a}(t) = z(t, 0)$, we set $\epsilon=0$:
\begin{equation}
    \label{result_int}
    \begin{aligned}
        \tilde{a}(t) &= a(t) \exp(-\gamma t) \\
         &+ \gamma \int_0^t \tilde{a}(t') g(t-t') \exp(-\gamma (t-t')) \mathrm{d} t'
    \end{aligned}
\end{equation}

\subsection{Expressing the solution in terms of Memory-Kernels}
In this section we use the laplace-transform to rewrite the result (\ref{result_int}) in terms of memory-kernels and to prove (\ref{main_res}).
Therefor we introduce the following laplace-transforms:
\begin{equation}
    \begin{aligned}
        a(t) \laplace A(s), \tilde{a}(t) \laplace \tilde{A}(s) \\
        K(t) \laplace \kappa(s), \tilde{K}(t) \laplace \tilde{\kappa}(s)
    \end{aligned}
\end{equation}

By transforming (\ref{mk_def}) for the decoupled case ($\gamma=0$), we find:
\begin{equation}
    \label{mk_def_laplace}
    \begin{aligned}
        s A(s) - a_0 = -\kappa(s) A(s)
    \end{aligned}
\end{equation}

$a(0) = \tilde{a}(0) = a_0$ follows from (\ref{result_int}).

We now prove that the dynamics $\tilde{a}(t)$ of the coupled system (\ref{result_int}) is generated by the 
memory-kernel $\tilde{K}(\tau)$, which is given by our main theorem (\ref{main_res}).



The laplace-transform of (\ref{main_res}) reads:
\begin{equation}
    \label{main_res_laplace}
    \tilde{\kappa}(s) = \kappa(s+\gamma)
\end{equation}

Firstly we transform (\ref{result_int}) and use (\ref{mk_def_laplace}):
\begin{equation}
    \label{zs1}
    \begin{aligned}
        \tilde{A}(s) =& A(s+\gamma) + \gamma \frac{1}{a_0} \tilde{A}(s) A(s+\gamma) \\
            =& (a_0+\gamma \tilde{A}(s)) \frac{1}{s+\gamma+\kappa(s+\gamma)} \\
            =& (a_0+\gamma \tilde{A}(s)) \frac{1}{s+\gamma+\tilde{\kappa}(s)}
    \end{aligned}
\end{equation}
In the last step we used (\ref{main_res_laplace}).

This equation can be transformed algebraically:
\begin{equation}
    \label{proof_goal}
        s \tilde{A}(s) - a_0 = -\tilde{\kappa}(s) \tilde{A}(s)
\end{equation}

By applying the inverse laplace-transform we finally find:
\begin{equation}
    \dot{\tilde{a}}(t) = -\int_0^t \tilde{K}(t-t') \tilde{a}(t') \mathrm{d}t'
\end{equation}
Thus the dynamics $\tilde{a}(t)$ of the coupled system is generated by the memory-kernel, which is given by our main theorem.

\end{document}